\documentclass[runningheads]{llncs}

\usepackage[T1]{fontenc}

\usepackage{graphicx}

\usepackage{amsmath}
\usepackage{amssymb}
\usepackage{booktabs}
\usepackage{xeCJK}
\usepackage{appendix}
\usepackage[ruled,vlined,linesnumbered]{algorithm2e}
\usepackage{xcolor}
\usepackage{xurl}
\usepackage{cite}
\usepackage[breaklinks=true]{hyperref}
\usepackage{float}
\begin{document}

\title{Array-Carrying Symbolic Execution for Function Contract Generation}

\author{Weijie Lu$^{*}$\inst{1}
\and
Jingyu Ke$^{*}$\inst{1}
\and
Hongfei Fu\inst{1}
\and
Zhouyue Sun\inst{1}
\and
Yi Zhou\inst{1}
\and
Guoqiang Li\inst{1}
\and
Haokun Li\inst{2}
}

\authorrunning{Weijie Lu et al.}

\institute{Shanghai Jiao Tong University, Shanghai, China \and
Peking University, Beijing, China
}

\maketitle

\begingroup
\renewcommand{\thefootnote}{}
\footnotetext{* These authors contributed equally to this work.}
\endgroup
\setcounter{footnote}{0}

\vspace{-6ex}
\begin{abstract}
Function contract generation is a classical problem in program analysis that targets the automated analysis of functions in a program with multiple procedures.
The problem is fundamental in interprocedural analysis where properties of functions are first obtained via the generation of function contracts and then the generated contracts are used as building blocks to analyze the whole program. Typical objectives in function contract generation include pre-/post-conditions and assigns information (that specifies the modification information over program variables and memory segments during function execution). In programs with array manipulations, a crucial point in function contract generation is the treatment of array segments that imposes challenges in inferring invariants and assigns information over such segments. To address this challenge, we propose a novel symbolic execution framework that carries invariants and assigns information over contiguous segments of arrays. We implement our framework as a prototype within LLVM, and further integrate our prototype with the ANSI/ISO C Specification Language (ACSL) assertion format and the Frama-C software verification platform. Experimental evaluation over a variety of benchmarks from the literature and functions from realistic libraries shows that our framework is capable of handling array manipulating functions that indeed involve the carry of array information and are beyond existing approaches.

\end{abstract}

\vspace{-6ex}
\section{Introduction}
\label{sec:intro}

In program analysis, functional contract generation aims at the automated generation of contracts that specify the input-output relationship of an individual function. The input-output relationship typically involves preconditions and postconditions, and may also include assigns information (from the ANSI/ISO C Specification Language (ACSL) assertion format) that specifies possible modifications over memory segments. The problem is fundamental in program analysis, as it facilitates interprocedural analysis by providing the basic information of functions. In this work, we focus on function contract generation for array-manipulating programs.

In the literature, various approaches have been proposed to address the automated generation of function contracts, including abstract interpretation~\cite{DBLP:conf/vmcai/CousotCFL13,DBLP:conf/vmcai/BoutonnetH19,invariant-abstract-interpretation,origin-abstract-interpretation,parametric-segmentation,DBLP:conf/sas/McCloskeyRS10,DBLP:conf/popl/GopanRS05}, model checking/SMT solving~\cite{DBLP:journals/fmsd/KomuravelliGC16,Array-BMC,Array-MC}, instrumentation~\cite{DBLP:conf/cav/AmilonEGLR23}, large language models (LLMs) ~\cite{WCSXQHLCT24,MLLXB25}, etc. Abstract interpretation chooses a suitable abstract domain for the abstraction of
program states, and then performs a top-down analysis for each function. Existing abstract interpretation approaches accommodate contiguous array memory segments and support the inference of array invariants ~\cite{invariant-abstract-interpretation,parametric-segmentation,DBLP:conf/popl/GopanRS05}. Model checking/SMT solving approaches~\cite{Array-BMC,Array-MC} encode program execution as rules and leverage SMT solvers to verify whether an assertion in the function holds or not. The treatment of contiguous memory segments can be readily encoded by SMT constraints. Instrumentation ~\cite{DBLP:conf/cav/AmilonEGLR23} allows a user to specify rules to transform programs into a form with ghost variables that is easier to analyze, where specific array patterns (e.g., summation, maximum)  are handled by special instrumentation rules.
LLM-based approaches (e.g.~\cite{WCSXQHLCT24,MLLXB25}) perform supervised fine-tuning (SFT) using manually annotated programs as training data. The resulting fine-tuned model then infers function contracts using suitable prompts, often with possible guidance from static analysis and external experts. 

While existing approaches tackle function contracts with array manipulations from different perspectives, they are limited in various aspects. For example, abstract interpretation has developed relational and segment-oriented domains that can reason about many array invariants~\cite{invariant-abstract-interpretation,parametric-segmentation,DBLP:conf/popl/GopanRS05,origin-abstract-interpretation,DBLP:conf/vmcai/BoutonnetH19}. However, for fully automated contract synthesis, preserving path-sensitive disjunctive loop-exit information and translating it into precise merged array-segment \texttt{assigns} information remains challenging.
Model checking/SMT solving approaches can also infer summaries/contracts in multiple settings~\cite{DBLP:journals/fmsd/KomuravelliGC16,Array-BMC,Array-MC,TriCera,Esen2022Tricera}. Nevertheless, depending on the encoding and property setup, many workflows are geared toward proving specific assertions, and may require additional summarization layers to produce reusable function-level contracts of the form \(\langle \textit{Pre}, \textit{Post}, \textit{Assign}\rangle\). Instrumentation focuses on manual specification of ghost variables, and hence does not have a dedicated treatment of contiguous array segments. LLM approaches usually require manual guidance from experts to avoid producing incorrect results caused by hallucinations. As a result, it remains a challenge to build a framework that handles array segments and invariants in function contract generation.

To address the aforementioned challenge, we propose a novel symbolic execution framework that carries array invariants and has a dedicated treatment of the contiguous array segments. Compared with previous results and existing symbolic execution techniques, our symbolic execution framework carries invariants generated from external invariant generators along the program statements, and carefully handles disjunction, splitting and merging of contiguous array segments. In our implementation, the invariant-generation back-end combines an affine/linear generator (the StInGX-style plugin based on Farkas-lemma reasoning~\cite{DInvG,StinGX}) with idiom-oriented generators for search and max/min loops. Appendix~\ref{app:generator-examples} gives concrete examples of these generators and the ACSL clauses they produce. The detailed contributions are as follows.

\begin{itemize}
\item Our symbolic execution is designed to receive numerical and array invariants from external invariant generators and propagate them along the execution of program statements. This allows our framework to leverage loop invariants to generate function contracts for general properties. Moreover, the propagation of invariants also produces tight preconditions for subsequent loop analysis.

\item Our symbolic execution has a refined treatment of contiguous array segments. In detail, our framework handles disjunction of array segments resulting from different program execution paths (including different loop exit modes),  and the merge/split of array segments during the symbolic execution. The merge and split of array segments allow precise tracking of information over these segments that is crucial to derive overall contracts related to array segments.

\item We implement our symbolic execution technique as a prototype that comprises about $19000$ lines in C++ within the LLVM platform~\cite{LLVM19,Lattner2004LLVM}.

We also integrate our prototype with the ACSL assertion format and the Frama-C software verification platform~\cite{FramaC31,Cuoq2012FramaC}, so that the result of our symbolic execution can be fully verified by external verifiers.
\end{itemize}
Experimental evaluation over a wide range of benchmarks (including realistic benchmarks from active libraries) shows that our framework is capable of generating contracts that describe precise behavior of array-manipulating functions, while previous approaches fail to handle many of these benchmarks.

\smallskip
\noindent{\em Limitations}. Our symbolic execution has limited ability to handle pointers. Currently, we can only handle simple pointer arithmetic with base and offset arising from array addresses, and cannot handle recursive pointer structures such as linked lists and binary trees.

\vspace{-2ex}
\subsection{Related Works}

Below we describe related works on function contract generation.

\smallskip
\noindent\emph{Abstract interpretation.} Abstract interpretation~\cite{origin-abstract-interpretation,abstract-interpretation-contracts} generates function contracts by a top-down analysis that propagates elements from an abstract domain along the execution of the program, where functions are inlined upon procedure calls. Existing abstract interpretation methods handle arrays by abstract domains that express invariants over contiguous memory segments of arrays~\cite{invariant-abstract-interpretation,parametric-segmentation}. As we consider to extend symbolic execution with array segments and invariants, our result has a different focus from abstract interpretation methods.

\smallskip
\noindent\emph{Model checking.} Model checking approaches~\cite{Array-BMC,Array-MC} often encode read/write operations of arrays by SMT rules and then solve target assertions with arrays. The most relevant works are by Tricera~\cite{TriCera} and a SeaHorn-based verifier\cite{SeaHorn-based} that generate function contracts through a transformation into CHC solving (via e.g. the state-of-the-art CHC solver Eldarica~\cite{Eldarica}). A major disadvantage of existing model checking approaches is that they often follow top-down analysis, so the generated specification is often highly specific to the initial values of a program. Our approach circumvents this limitation by leveraging external invariant generators in our symbolic execution.

\smallskip
\noindent\emph{Large language models.} In recent years, there has been a burgeoning focus on applying Large Language Models (LLMs) to support code specification generation~\cite{KGVJ24,XYJKTZL25,LLM25,MLLXB25}. A standard framework is AutoSpec~\cite{WCSXQHLCT24} that leverages LLMs to synthesize ACSL-language specifications for C programs via a static analysis-guided bottom-up iterative approach, offering wider applicability than traditional works. Additionally, further studies~\cite{PBSSY23,PRBC24,LQFCY24,KKK25,LCWKL26} mainly focus on LLM-based invariant generation.

\smallskip
\noindent{\em Symbolic execution.}
Existing symbolic execution techniques analyze program with memory operations by symbolically encoding program states and maintaining a symbolic memory model that tracks heap and array updates across execution paths. For example,
\textsc{Clang Static Analyzer}~\cite{Clang} performs a path-sensitive symbolic execution with a region-based memory model, but its reasoning over arrays and heap is local and non-relational. It propagates per-element constraints along each path, detecting out-of-bounds or null dereferences, yet lacks the ability to summarize iteration effects or infer relational properties over contiguous array regions across loop executions.
\textsc{KLEE}~\cite{KLEE} symbolically executes LLVM bitcode with SMT-based constraint solving, representing memory as byte-addressable symbolic objects and focusing on path exploration rather than invariant inference.
While several extensions~\cite{UCKLEE,Seg-KLEE} enhance its handling of heap-allocated objects and symbolic pointers, KLEE and its variants still analyze memory at the level of individual allocations and lack the capability to infer relational or inductive invariants over contiguous array regions.
The Gillian platform~\cite{GillianPartI2020,GillianPartII2021} provides a language-parametric symbolic execution architecture, and further integrates separation-logic-based verification with compositional memory models for JavaScript and C. Compared with Gillian, our work focuses on automatic contract synthesis for array-manipulating functions in C, where quantified array invariants are carried and merged along symbolic execution to synthesize ACSL \texttt{requires}/\texttt{ensures}/\texttt{assigns} clauses.
Compared with existing results, our approach extends symbolic execution with contiguous memory segments of arrays.

\smallskip
\noindent\emph{Invariant generation.}
Invariant generation
can be addressed by abstract interpretation ~\cite{invariant-abstract-interpretation,parametric-segmentation,DBLP:conf/popl/GopanRS05,origin-abstract-interpretation,DBLP:conf/vmcai/BoutonnetH19}, constraint solving~\cite{DBLP:conf/vmcai/LarrazRR13,DBLP:conf/cav/ColonSS03,DInvG,DBLP:conf/pldi/Chatterjee0GG20,StinGX}, instrumentation~\cite{DBLP:conf/cav/AmilonEGLR23},
recurrence analysis~\cite{Non-linear-invariant,DBLP:journals/pacmpl/WangL23}, etc.
Low-level Bi-Abduction~\cite{LowLevelBiAbduction2022} is particularly relevant for compositional analysis of low-level pointer-intensive fragments, inferring function contracts through separation-logic bi-abduction with support for operations such as pointer arithmetic and block splitting/merging. In contrast, our target is contiguous array reasoning with carried quantified invariants and disjunctive segment summaries, which is tailored to postcondition and assigns synthesis in ACSL rather than low-level heap-shape reasoning over intrusive linked structures.
Moreover, by integrating a Linear/Affine invariant generator plugin (in the style of StInGX/Farkas-lemma-based inference~\cite{StinGX,DInvG}) into our loop-analysis pipeline, our framework can effectively handle a broad class of numerical loops as well, including loops whose key properties are captured by affine index/value relations. Our approach utilizes these methods to generate numerical and array invariants to be carried during our symbolic execution.

\section{Overview}
\label{overview}

Below we have an overview of our results via two illustrative examples.

\begin{figure}[htbp]
    \centering
    \includegraphics[width=1\textwidth]{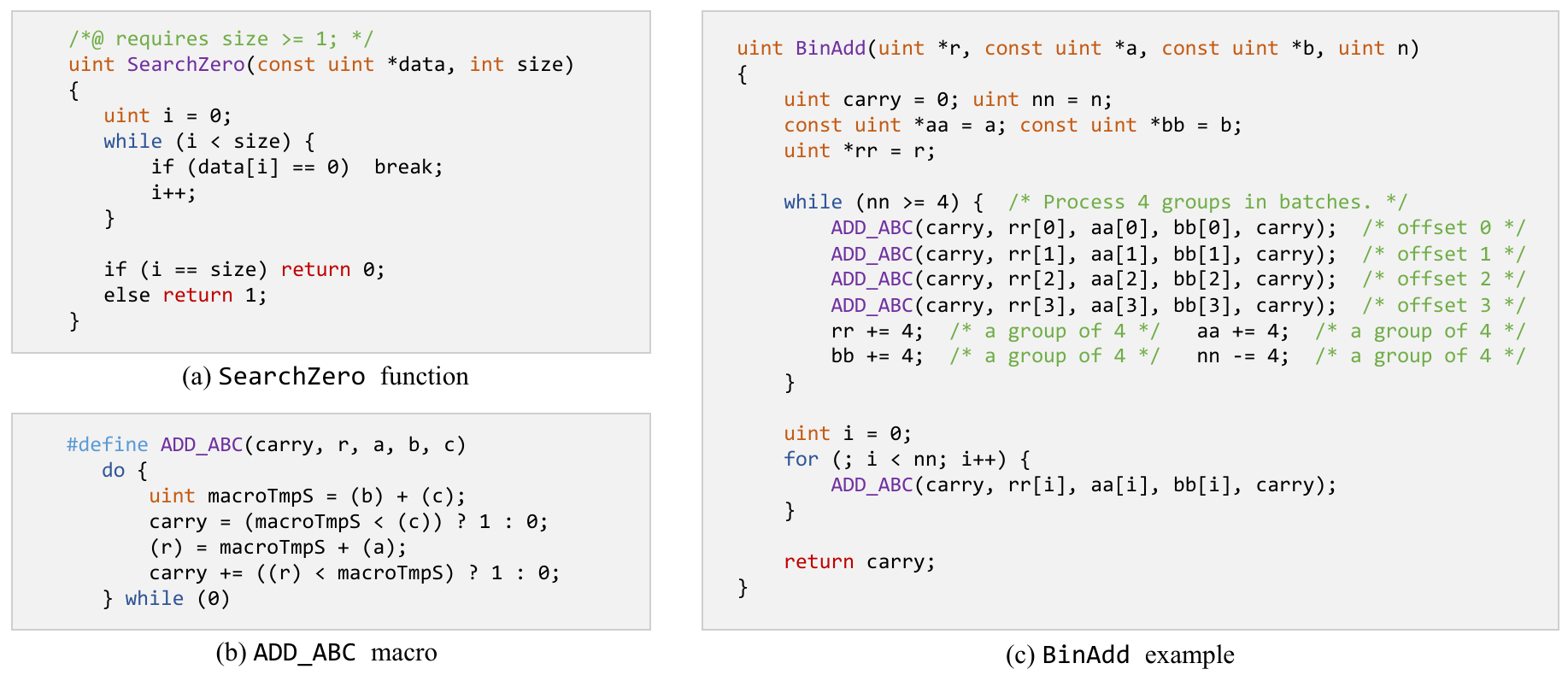}
    \caption{Array program examples}
    \label{fig:code_example}
\end{figure}

\smallskip
\noindent\emph{First example.} Consider the array manipulating function shown in Figure~\ref{fig:code_example}(a), which is a function that searches whether there is a zero entry in the array $\mathsf{data}$. The precondition of the function is $\mathsf{size}\ge 1$ which specifies that the size of the array should be positive. A notable feature of this function is that after the while loop, the function continues to check whether $i$ equals $\mathsf{size}$. If so, then the function returns 0, meaning that no zero entry is found.
Otherwise, the function returns 1, meaning that a zero entry has been found.

To generate a complete contract for this function, our symbolic execution maintains both array invariants and contiguous array segments. For this example, our symbolic execution first calls an external invariant generator to generate loop invariant and summary for the while loop. By distinguishing different exit ways of the loop,
our symbolic execution receives a disjunctive loop summary of two branching execution paths: one corresponds to the case when the loop exits upon encountering the conditional branch \texttt{data[i] != 0}, denoted as \(\pi_0\), and the other corresponds to the case when the condition \texttt{data[i] == 0} consistently holds, denoted as \(\pi_1\). Based on these execution paths, we extract the loop summary with information over the contiguous array segment with index from $0$ to $i$ as follows.
Specifically, along the path \(\pi_0\),
our symbolic execution receives the assertion at the loop exit: $(\forall j.\ 0 \le j < i \rightarrow \mathrm{data}[j] = 0)
\ \land\
\mathrm{data}[i] \ne 0
\ \land\
i < \mathrm{size}$. On the other hand, along the path \(\pi_1\), our symbolic execution receives $(\forall j.\ 0 \le j < i \rightarrow \mathrm{data}[j] = 0)
\ \land\
i = \mathrm{size}$. Note that the assertions of the two paths cannot be merged as they differ at whether \texttt{data[i] == 0} and \texttt{i == size}.  Our symbolic execution maintains the disjunction of these two assertions, and carries it forward to the remaining part of the program so that the function contract below
\[
\begin{aligned}
\textbf{Pre}:&\ \text{size} \ge 1 \qquad \textbf{Assigns}:\ \varnothing\\
\textbf{Post}:&\
\bigl[(i=\mathrm{size}) \land (\forall j.\ 0 \le j < \mathrm{size} \rightarrow \mathrm{data}[j]=0) \wedge \text{res} = 0\bigr] \\
&\ \lor\
\bigl[(i<\mathrm{size}) \land (\mathrm{data}[i]\neq 0) \land (\forall j.\ 0 \le j < i \rightarrow \mathrm{data}[j]=0) \wedge \text{res} = 1\bigr]. \\
\end{aligned}
\]
is inferred, capturing precisely the behavior of the function. The main contribution of our symbolic execution here is the carry of (disjunctive) invariants and contiguous array segments.

\smallskip
\noindent
\emph{Second example.} Consider the BnADD function extracted from the openHiTLS library \cite{openhitls}, as shown in Figure~\ref{fig:code_example}(c)  (with macro given in Figure~\ref{fig:code_example}(b)). To capture the overall memory modification in the execution of the function,
one needs to merge the range of modified array segments in the two loops. Our symbolic execution detects
changed array segments in each loop of the function, and merge the two segments into a single segment.
In this example, the modified array range of the first loop is detected as \(\mathrm{rr}[0..4k)\), corresponding to the batched updates of four elements per iteration, where \(k\) is a ghost variable implicitly defined within the loop to represent the number of iterations executed so far.
Informally, the term \(\mathrm{rr}[0..4k)\) means that the content from index $0$ to index $4k$ of the array $\mathrm{rr}$ may be modified after $k$ loop iterations.
The counterpart of the second loop is detected as \(\mathrm{rr}[4k..\mathrm{n})\), which handles the remaining elements after the batch processing.
By merging the two modified ranges, our symbolic execution generates the overall modified range \(\mathrm{rr}[0..\mathrm{n})\).
The major contribution here is the merging of modified memory segments in our symbolic execution that allows us to generate the complete assigns information in the function contract.

\section{Array-Carrying Symbolic Execution}
\label{sec:array symbolic exec}

In this section, we illustrate our symbolic execution framework over functions with integer and array variables that carries invariants and  contiguous array segments, and aims at automatically synthesizing a function contract for a function in a given C program.

Due to the space constraints, we present the key concepts of invariants and function contracts in our symbolic execution informally, and relegate the detailed definitions (over a simplified program model in Appendix~\ref{sect:progmodel}) to Appendix~\ref{sect:invs and contract}. An \textit{invariant} is a collection of formulas at certain program counters required to hold in every program state along every feasible execution path, thereby over-approximating the set of reachable program states. A \textit{function contract} characterizes the intended behavior of a function by a \textit{precondition} restricting admissible input program states, a \textit{postcondition} describing the property that must hold upon termination, and an \textit{assigns} information indicating possible memory locations whose content may be modified during function execution.

\begin{figure}[htbp]
    \centering
    \includegraphics[width=1\textwidth]{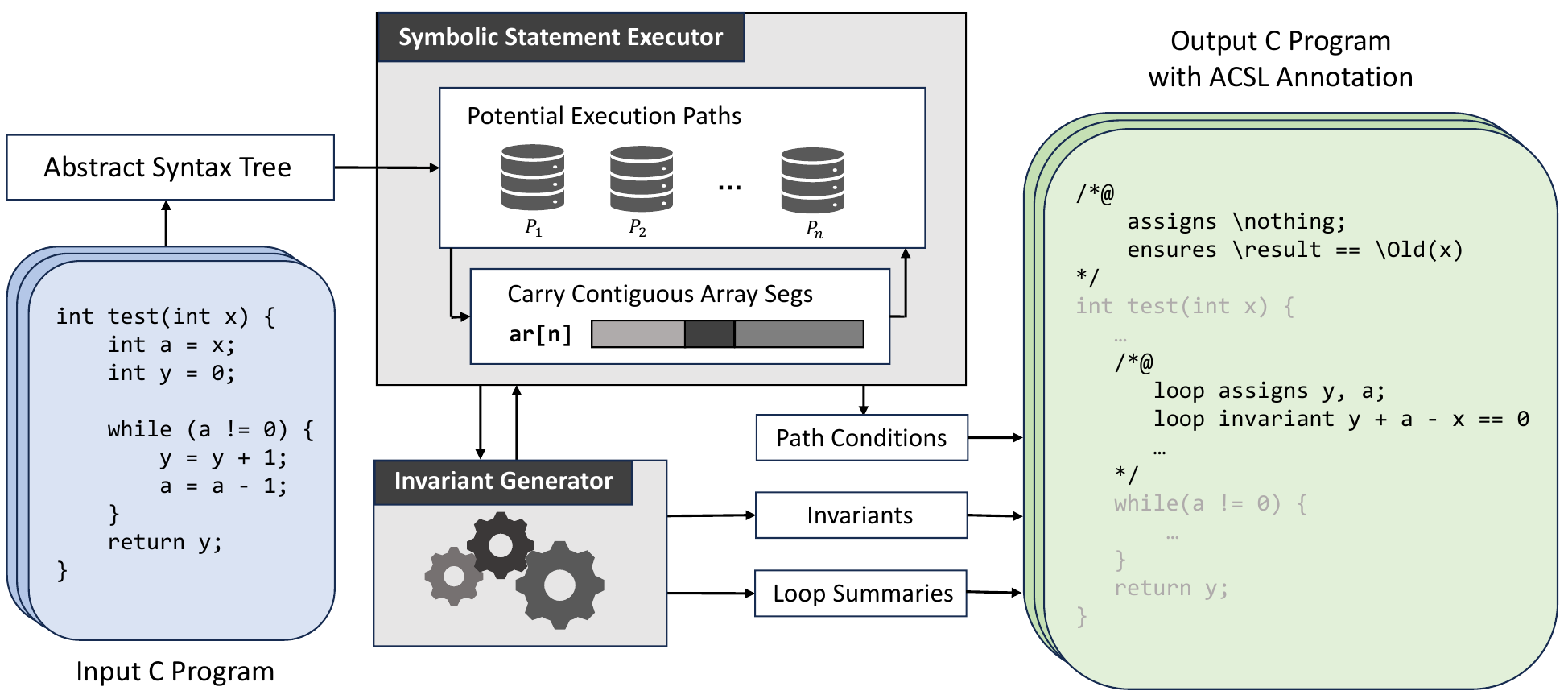}
    \caption{General workflow of array-carrying symbolic execution}
    \label{fig:workflow}
\end{figure}

\smallskip
\noindent{\em Overview.}
Figure~\ref{fig:workflow} demonstrates the workflow of our symbolic execution framework. The framework first takes as input an abstract syntax tree of the function body of the function to be analyzed. Then, a symbolic statement executor simulates the execution of the function body, producing multiple potential execution paths. During the simulation, the statement executor calls external invariant generators to generate invariants and summaries for loops. Moreover, the executor carefully maintains contiguous array segments of array contents and invariants, and carries them forward in the remaining execution. Finally, the execution paths with their path conditions, together with the carried invariants and contiguous array contents, are combined to obtain the function contract of the whole function. To maintain array segments and invariants, our symbolic statement executor incorporates an extended memory model to support constraint inference involving arrays and pointers.

Below we first present the high-level workflow of our framework, then the technical details on handling invariants and contiguous array segments, and finally a prototype implementation for our framework in LLVM.

\vspace{-1ex}
\subsection{Symbolic Execution Framework}
\label{subsec:symbolic exec framework}

In this section, we introduce our symbolic execution framework. We focus on the read and write aspects of arrays, and omit other technical aspects such as records and pointer arithmetic. A high-level program model is relegated to Appendix~\ref{sect:progmodel} for space constraints.

As our symbolic execution handles array operations, we model array access and modification operations by direct read and write operations at memory addresses of array variables. In our framework, we consider arithmetic expressions $\gamma$ with memory read operations by
$\gamma ::= c \mid x \mid a \mid \gamma_1 \oplus \gamma_2 \mid R(a)$. In the syntax, we have integer variables $x$, integer constants $c$, address variables $a$ and arithmetic operations $\oplus$ including addition, subtraction and multiplication (we omit division to avoid floating point errors and division by zero for simplicity).
Moreover, the expression $R(a)$ represents the result of the read operation at the address variable $a$.
Integer variables and constants are extracted from the underlying program, while address
variables are fresh variables referring to pointers. 

Let $\mbox{\sl Addr}$ be a countable set of \emph{address constants} and $V$ be a countable set of \emph{symbolic values} (for undetermined values).  We denote the counterpart of arithmetic expressions $\gamma$ where the role of program variables switches to symbolic values as \emph{symbolic expressions}, denoted by $\beta$ with possible subscripts. 

A \emph{symbolic address} is an ordered pair \((\beta_1,\beta_2)\) of symbolic expressions in the syntax above for which \(\beta_1\) is the \emph{base} expression and \(\beta_2\) is the \emph{offset} expression, which implicitly corresponds to the sum of the base and the offset (multiplied with the size of array entry). However, for the sake of clearly representing pointer arithmetic and intra-array offsets, we shall consistently use the pair notation \((\beta_1,\beta_2)\) throughout the remainder of the paper to denote all address values.
In our symbolic execution, we consider an intermediate program syntax as follows:
\begin{eqnarray*}
S ::= x := \gamma  \mid a:= \&x \mid W(a, \gamma)
\mid \text{if } b \text{ then } S_1 \text{ else } S_2
\mid \text{while } b \text{ do } S
\mid S_1 ; S_2~.
\end{eqnarray*}
In the grammar, we have assignment statements for integer variables $x := \gamma$, conditional branches $\text{if } b \text{ then } S_1 \text{ else } S_2$, while loops $\text{while } b \text{ do } S$, in which $b$ represents the boolean expressions built from comparison
of symbolic expressions and propositional logic connectives. In addition, we have a statement form $a:= \&x$ that assigns the symbolic address of $x$ to an address variable $a$,
corresponding to the address-of operation,
and a statement form $W(a, \gamma)$ that writes the expression $\gamma$ to the symbolic address of the address variable $a$. 

Our symbolic execution works with symbolic states. Formally, a \emph{symbolic state} $\Sigma$ is an ordered pair $(M,A)$ where $M$ is a \emph{memory map} that maps every symbolic address encountered so far to the symbolic expression held at the symbolic address, and $A$ is an \emph{address map} that maps each address and integer variable to its corresponding symbolic address where its concrete value lies. A symbolic state records the current symbolic addresses of the integer and array variables, and the current symbolic expressions
at the symbolic addresses. 

In our framework, read and write operations on addresses are realized by updating the corresponding maps in the current symbolic state. For example, the read expression \(R(a)\) for a symbolic state $(M,A)$ denotes the symbolic address $M(A(a))$ held in the address variable \(a\). 

The workflow of our symbolic execution is a traversal of program statements in the adapted program syntax $S$. During the traversal, we maintain a symbolic configuration $\mbox{\sl Conf}$. Formally, a \textit{symbolic configuration} $\mbox{\sl Conf}$ is a finite set of \emph{symbolic paths}, where each symbolic path $P\in \mbox {\sl Conf}$ is a tuple $(\ell, \Sigma, \mbox{\sl Seg},
\mbox{\sl pathCond})$ that specifies the current status after the execution of the path:
$\ell$ is the current program counter, $\Sigma$ is the current symbolic state, $\mbox{\sl Seg}$ is a structure that records the current information for contiguous array segments, and $\mbox{\sl pathCond}$ is a collection of path conditions resulting from the execution of the symbolic path. Invariants are treated as path conditions in $\mbox{\sl pathCond}$. 

Algorithm~\ref{alg:symexec} presents the pseudocode of our queue-based
symbolic execution procedure.
The main workflow is an enumeration of all feasible execution paths under a program in the syntax of $S$.
In this pseudocode, we require an input \texttt{PreCond} that corresponds to the input ACSL precondition (i.e., the \texttt{requires} clause of the target function). Here we focus on the update of $\ell,\Sigma$ and conditions in  $\mbox{\sl pathCond}$ along the symbolic execution, and relegate the details for the update of $\mbox{\sl Seg}$ and invariants in
$\mbox{\sl pathCond}$ to the next subsection.

The procedure begins by initializing the initial symbolic state $\Sigma_0=(M_0,A_0)$ at the entry of a program, as shown in Algorithm \ref{alg:symexec}, starting from line \ref{line:init}.
The initialization works as follows.

\begin{itemize}
    \item In the initial address map \(A_0\), each program variable \(x \in X\)
    is mapped to a symbolic address \(\mathrm{addr}(x)\) drawn from the set
    $\mbox{\sl Addr}$. As defined earlier, every \(\mathrm{addr}(x)\) takes the form
    \((\beta_1, \beta_2)\). The base address \(\beta_1\) is a fresh address constant; for
    integer variables and pointers, the offset address \(\beta_2\) is set to \(0\),
    whereas for an array element \(arr[i]\), the offset is the
    symbolic expression $M_0(A_0(i))$.

    \item For each symbolic address \(\mathrm{addr}(x)\) associated with a variable
    \(x\) in \(A_0\), we insert into \(M_0\) a mapping from \(\mathrm{addr}(x)\)
    to an initial symbolic expression \(\beta_x\). The expression \(\beta_x\)
    corresponds to the initial value assigned to \(x\) in the program. If such a
    value does not exist or cannot be statically determined, we instead assign a fresh symbolic value taken from $V$ as the \(\beta_x\). We also substitute program variables $x$ in the given precondition by
    $\beta_x$.

\end{itemize}

Throughout this initialization procedure, all symbolic addresses
\(\mathrm{addr}(x)\) and symbolic values used for different variables
are guaranteed to be distinct.

After the initialization, as shown in Algorithm \ref{alg:symexec} starting from line \ref{line:traversal}, the symbolic executor traverses the statements in the function.
Suppose that $S$ is the current statement in the traversal.
To symbolically execute $S$, the algorithm dequeues existing symbolic paths from the path queue, and for each path $P$, invokes the \texttt{Exec} procedure.
The \texttt{Exec} procedure serves as the \textit{symbolic transfer function}: it interprets the semantics of $S$ to compute the successor symbolic state and update the path conditions.
Depending on the specific type of the statement $S$, the symbolic path $P=(\Sigma=(M,A), \mbox{\sl pathCond})$ is updated by \texttt{Exec} as follows.
(Here we omit the components $\ell,\mbox{\sl Seg}$ in the path description as (i) the update of $\ell$ corresponds to the natural control flow successor of $S$ and (ii) the update of $\mbox{\sl Seg}$ will be detailed in the next subsection.)
Below, given a memory map $M$, we denote by $M[(\beta_1,\beta_2)\mapsto \beta]$ the memory map that maps the symbolic address $(\beta_1,\beta_2)$ to the symbolic expression $\beta$, while maintaining the existing mappings for all other symbolic addresses.

\begin{algorithm}[!t]
\small
\caption{Symbolic Execution and Contract Synthesis Workflow}
\label{alg:symexec}
\KwIn{\(Func\) as target function context, \(PreCond\) as the set of preconditions}
\KwOut{\(Conf, \mathcal{I}_{\mathit{all}}, \mathcal{C}_{\mathit{func}}\) as symbolic configuration, loop invariants and synthesized function contracts}

\BlankLine

Extract variable set \(X\) from \(Func\)\;\label{line:init}
\ForEach{\(x \in X\)}{
    \(A_0(x) \gets \mathrm{addr}(x)\);\quad \(\texttt{Init}(M_0(addr(x))\)\;
}
\(\Sigma_0 = (M_0, A_0)\);\ \(\mathit{pathCond}_0 \gets \varnothing\)\;
\(P_0 = (\Sigma_0, \mathit{pathCond}_0)\);\ \(\mathrm{Path}\).\texttt{enqueue}(\(P_0)\);\ \(\mathcal{I}_{\mathit{all}} \gets \varnothing\)\;

\BlankLine
\ForEach{\(S \in Func\) \text{in program order}}{\label{line:traversal}
    \(\mathrm{Path}'\gets \phi\)\;
    \While{\(Path\) is not empty}{
        Dequeue \(P = (\Sigma, \mathit{pathCond})\) from \(Path\)\;

        \uIf{\(S\) is \texttt{ifstmt}}{\label{line:ifstmt}
            \(P_{1} \gets (\Sigma,\, \mathit{pathCond} \land [b]_{\Sigma})\)\;
            \((\Sigma_{1}', \mathit{pathCond}_{1}') \gets \texttt{Exec}(S_1, P_{1})\)\;
            \(Path'\).\texttt{enqueue} \((\Sigma_{1}', \mathit{pathCond}_{1}')\)\;

            \(P_{2} \gets (\Sigma,\, \mathit{pathCond} \land [\lnot b]_{\Sigma})\)\;
            \((\Sigma_{2}', \mathit{pathCond}_{2}') \gets \texttt{Exec}(S_2, P_{2})\)\;
            \(Path'\).\texttt{enqueue} \((\Sigma_{2}', \mathit{pathCond}_{2}')\)\;
        }

        \uElseIf{\(S\) is \texttt{whilestmt}}{\label{line:whilestmt}
            \((M_{\text{loop}}, PC_{\text{loop}}, Inv) \gets \texttt{LoopAnalyzer}(S, P)\)\;
            \(\mathcal{I}_{\mathit{all}} \gets \mathcal{I}_{\mathit{all}} \cup Inv\)\;
            \((\Sigma', \mathit{pathCond}') = \texttt{Update}((M_{\text{loop}}, PC_{\text{loop}}, Inv), P)\)\;
            \(Path'\).\texttt{enqueue}\((\Sigma', \mathit{pathCond}')\)\;
        }

        \Else{ \label{line:assign} \tcp{Handle assignment statements}
            \((\Sigma', \mathit{pathCond}') \gets \texttt{Exec}(S, (\Sigma, \mathit{pathCond}))\);\;
            \(Path'\).\texttt{enqueue}\((\Sigma', \mathit{pathCond}')\)\;
        }
    }
    \(Path \gets Path'\)\;
}

\BlankLine

\(Conf \gets Path\);\ \(\mathcal{C}_{\mathit{func}} \gets \texttt{ContractGen}(Conf)\)\;
\Return \((Conf, \mathcal{I}_{\mathit{all}}, \mathcal{C}_{\mathit{func}})\)\;
\end{algorithm}

\begin{itemize}
\item If $S$ is a conditional statement with condition $b$, as shown in Algorithm \ref{alg:symexec} starting from line \ref{line:ifstmt}, then the symbolic executor splits the symbolic path $P$ into two new symbolic paths $P_b, P_{\neg b}$. The symbolic path $P_b$ equals $(\Sigma, \allowbreak \mbox{\sl pathCond} \cup \{[b]_\Sigma\})$, for which the condition  $[b]_\Sigma$ is obtained by replacing every integer program variable $x$ and read operation $R(a)$ in $b$ by their corresponding symbolic expressions $M(A(x))$ and $M(A(a))$, respectively.

The case for $P_{\neg b}$ is similar.

\item If $S$ is an assignment statement, as shown in Algorithm \ref{alg:symexec} starting from line \ref{line:assign}, then the symbolic path is updated to $(\Sigma', \mbox{\sl pathCond})$ for which
the updated symbolic state $\Sigma'$ depends on the type of the assignment.

\begin{enumerate}
    \item If $S$ is a variable assignment $x := \gamma$, we evaluate $\gamma$ under $\Sigma$ (by substituting each integer or address variable $u$ into symbolic expressions stored at $A(u)$, namely $M(A(u))$, and each symbolic address $(\beta_1,\beta_2)$ into the symbolic expression $M(\beta_1,\beta_2)$ held at the address)
    to obtain a symbolic expression $[\gamma]_\Sigma$, and then update the memory map at the address of $x$: $\Sigma' = (\, M[ A(x) \mapsto [\gamma]_\Sigma ],\; A \,)$.

    \item If $S$ is an address-of statement $a := \&x$, we update the symbolic expression stored at $A(a)$ in the memory map with the address value $A(x)$: $\Sigma' = (\, M[ A(a) \mapsto A(x) ],\; A \,)$.

    \item If $S$ is a write-back statement $W(a, \gamma)$, we evaluate $\gamma$ to obtain $[\gamma]_{\Sigma}$, as described earlier. Then we update the corresponding memory entry at $M(A(a))$, i.e. the memory location at the address expression stored in $a$:  $\Sigma' = (\, M[M(A(a))\mapsto [\gamma]_{\Sigma} ],\; A \,)$.
\end{enumerate}

\item If the next statement is the entry of a while loop, as shown in Algorithm \ref{alg:symexec} starting from line \ref{line:whilestmt}, the symbolic executor invokes an external loop analyzer. For each symbolic path in the current configuration, the executor sends the current symbolic state and path conditions, as the precondition for the loop), to the analyzer, which returns a set of inferred invariants and a loop summary.

Let $M_{\text{loop}}$ denote the modifications applied to the memory map during the loop execution, $PC_{\text{loop}}$ the path conditions of the loop, and $Inv$ the set of inferred loop invariants. The invariants $Inv$ are added to the global collection $\mathcal{I}_{\mathit{all}}$ for contract synthesis. The executor then transfers
control to $PC_{\mathit{loop}}$ and updates the symbolic path accordingly. This
update has two components: (1) conjoining the inferred invariants with the
current path condition, and (2) updating the memory map to incorporate the loop
effects. For post-loop variable states, the executor introduces fresh symbol values for the exit values of the program variables to facilitate subsequent symbolic execution

when
necessary—namely, for variables modified inside the loop whose final values
cannot be precisely determined or summarized. Our current implementation uses a
preliminary loop analyzer described in Appendix~\ref{sect:loopanalysis}.

\end{itemize}

After the symbolic execution, the overall function contract $\mathcal{C}_{\mathit{func}}$ is inferred
from the final symbolic configuration \(Conf\).
The \texttt{assigns} clause is derived by comparing the final memory map $M$ of each path against the initial memory map $M_0$ from the function entry: any memory location whose final symbolic value differs from its initial value is recorded as modified. The logical components of the contract are then extracted from the accumulated
path conditions \mbox{\sl pathCond} and the segment structural information in
\(\mbox{\sl Seg}\). The projection step is performed by eliminating intermediate
symbols introduced during execution and retaining only symbols associated with
the function entry (for \texttt{requires}) and return state (for
\texttt{ensures}/\texttt{assigns}). Concretely, we substitute each symbolic
address with the corresponding source-level variable term, rewrite segment
constraints into index-range predicates over arrays, and existentially quantify
auxiliary symbols that do not correspond to user-visible program variables.
The resulting formulas characterize the preconditions and postconditions under
which the function is well-defined and the properties that must hold of its outcomes.

\vspace{-3ex}
\subsection{The Carrying of Invariants and Contiguous Array Segments}
\label{subsec:invariants and array segs}

\smallskip
\noindent\emph{Invariant Carrying.}
The path-condition set $\mbox{\sl pathCond}$ records constraints over the symbolic expressions containing symbolic values which are generated during execution, rather than the program variables that may temporarily reference those expressions. When a loop invariant and its corresponding summary are obtained from the external invariant generator, the symbolic executor may introduce fresh symbolic values from $V$ to represent the values of the integer variables upon loop termination. The loop summary is then instantiated by substituting these fresh symbolic values for the concrete integer variables, and the resulting condition is appended to $\mbox{\sl pathCond}$. The symbolic execution of the remaining program proceeds using these newly introduced symbolic values as the initial values for their respective integer variables.

\smallskip
\noindent\emph{Representing Contiguous Array Segments.}
We formally define contiguous array segments using our address model.
As mentioned earlier, each symbolic path carries a set of active segments, denoted as $\mbox{\sl Seg}$. Unlike the memory map $M$ which can only record independent symbolic expressions stored at discrete symbolic addresses, segments capture the relationship between the properties of a contiguous memory region and the symbolic expressions stored over that region. Based on this, one can derive different patterns of  symbolic expressions that serve as constraint conditions associated with that memory segment. 

We define a \textit{segment} $s \in \mbox{\sl Seg}$ as a triple $s := \langle (\beta_{b}, \beta_{o}),\; \beta_{len},\; \mathcal{C} \rangle$, where $(\beta_{b}, \beta_{o})$ denotes the starting symbolic address of the segment, $\beta_{len}$ is a symbolic expression representing the number of elements in the segment, and $\mathcal{C}$ denotes the \emph{content summary} of the segment, representing the pattern of invariants that hold over this segment.

Our framework employs a hybrid strategy to obtain the content summary $\mathcal{C}$:

\begin{enumerate}
    \item \emph{Directly-Indexed Invariants:}
    When the array content can be expressed as a function of the index, $\mathcal{C}$ is a symbolic expression $\phi(i_{off})$. Here, $\phi$ is a function representing the invariant pattern, and $i_{off}$ is a placeholder variable representing the relative offset within the segment. For example, given the constraint $\forall 0 \leq i \leq n-1, {\mbox{\sl arr}}[i] = i + 2$ over the array $\mbox{\sl arr}$, where $n$ denotes the length of $\mbox{\sl arr}$, one can derive the segment $s = \langle (A(arr), 0), n-1, \phi(i_{off}):=i_{off}+2\rangle$.

    \item \emph{Aggregate Abstractions:}
    Certain program logic relies on reduction operations~\cite{DBLP:conf/cav/AmilonEGLR23} (e.g., \texttt{sum}, \texttt{max}) over an array range.
    In such cases, we model $\mathcal{C}$ as a first-class arithmetic expression, denoted as $\beta_{agg} := \texttt{Agg}(\oplus, s)$.
    Here, $\oplus$ represents the reduction operator and $s$ is the target segment.
    Unlike logical constraints which are restricted to $\mbox{\sl pathCond}$, these aggregate expressions behave identically to standard symbolic expressions.
    Consequently, they can be inserted into the the memory map $M$ as the symbolic expressions associated with the corresponding symbolic addresses and participate in subsequent arithmetic computations, enabling algebraic reasoning over array properties.
\end{enumerate}

    Certain loop-based properties require quantification over the actual iteration
    domain of the loop (i.e., the iterator $k$ ranging over the loop bounds).
    Such invariants cannot be captured through simple index-pattern abstractions
    and therefore are not stored inside the segment tuple.
    Instead, they are represented explicitly as logical statements and added to
    the path condition set $\mbox{\sl pathCond}$.

    We formalize these complex invariants as \textit{quantified constraints}.
    Formally, such a constraint is a tuple $\Phi := \langle \mathcal{Q}, \mathcal{R}, \psi \rangle$, representing the logical formula $\mathcal{Q}\, k \in \mathcal{R} . \psi(k)$, where $\mathcal{Q} \in \{ \forall, \exists \}$ is the quantifier, $\mathcal{R}$ denotes the index range, typically defined by bounds $[\beta_{lower}, \beta_{upper})$, and $\psi(k)$ is the predicate over the loop iterator $k$. For instance, consider the \texttt{SearchZero} function presented in the overview, shown in Figure~\ref{fig:code_example}(a).
    The loop iterates through the array to find the first non-zero element, effectively maintaining the invariant that all elements scanned so far are zero.
    Our framework captures this property as the tuple $\langle \forall, [0, i), \texttt{data}[k] = 0 \rangle$.
    Here, the quantifier $\forall$ applies to the index range $[0, i)$ (where $i$ is the current loop variable), and the predicate $\texttt{data}[k] = 0$ asserts the specific constraint validated by the loop logic.

\smallskip

\noindent\emph{Updates on Segments and Invariants.}
During symbolic execution, the executor maintains and forwards the array-related information encoded in the segments, and performs the corresponding segment updates whenever a related array memory operation is executed.

\begin{itemize}

    \item \emph{Read $R(a)$:}
    Let $a = (\beta_{arr}, \beta_{len})$ be the symbolic address being read, where the offset address
    $\beta_{len}$ is a symbolic expression evaluated under the current symbolic state. The executor first locates all segments
    $s = \langle (\beta_{arr}, \beta_o), \beta_l, \mathcal{C} \rangle \in \mbox{\sl Seg}$
    whose index range may overlap with $\beta_{len}$, i.e., all $s$ satisfying
    \(\beta_o \le \beta_{len} < \beta_o+\beta_l\).
    For each such segment, the executor performs an index-based substitution,
    replacing the placeholder \(i_{idx}\) in its content summary $\mathcal{C}$
    with the offset \(\beta_{len}\), yielding the expression $\mathcal{C}(\beta_{len})$. If multiple segments are simultaneously feasible (e.g., due to symbolic index
    uncertainty), the executor constructs a disjunctive invariant of the form $\bigvee_{\mathcal{C} \in \mathsf{Cand}} \mathcal{C}(\beta_{len})$,
    where $\mathsf{Cand}$ is the set of all feasible segments.

    \item \emph{Write $W(a, \gamma)$:} A write operation $W(a, \gamma)$ to an address $a$ falling within a segment $s$ may invalidate or split the segment.
    If $a$ represents a concrete offset relative to the segment base, $s$ is split into two or more new, smaller segments.
    For instance, writing to \texttt{arr[j]} may split a target segment $s = \langle (\beta_{arr}, 0), n, \mathcal{C} \rangle$ into a \textit{prefix segment} $s_1 = \langle (\beta_p, 0), j, \mathcal{C}_1 \rangle$ and a \textit{suffix segment} $s_2 = \langle (\beta_p, j+1), n-j-1, \mathcal{C}_2 \rangle$.
    The original content summary $\mathcal{C}$ is propagated to these new, smaller segments.
    In the current prototype, this splitting operation is implemented for segments with constant offset and length, while segments with symbolic offset or length remain a subject for future work.

    \item \emph{Segment Merging:}
    The framework also merges segments to maintain coarse-grained array summaries.
    After a write operation or loop analysis, a syntactic merging strategy is applied.
    This procedure scans all segments sharing the same base address and identifies pairs of adjacent segments---e.g.,
    $s_1 = \langle (\beta_{b}, \beta_{o}), \beta_{L1}, \mathcal{C}_1 \rangle$ and
    $s_2 = \langle (\beta_{b}, \beta_{o} + \beta_{L1}), \beta_{L2}, \mathcal{C}_2 \rangle$.
    If their content summaries are syntactically equivalent, the two are merged into a single, larger segment
    $s_{\mathit{new}} = \langle (\beta_{b}, \beta_{o}), \beta_{L1} + \beta_{L2}, \mathcal{C}_1 \rangle$.

\end{itemize}

We provide an example for the carrying process as follows.

\begin{example}
Consider the first code example in
Figure~\ref{fig:code_example}(a). Starting from the input ACSL precondition
\(\mathsf{size}\ge 1\), symbolic execution initializes \(i\) and the symbolic
state, then invokes the loop analyzer at the \texttt{while} entry. The analyzer
returns a disjunctive summary: one branch exits with
\((\forall 0\le j<i.\ \mathsf{data}[j]=0) \land \mathsf{data}[i]\neq 0 \land
i<\mathsf{size}\), while the other exits with
\((\forall 0\le j<i.\ \mathsf{data}[j]=0)\land i=\mathsf{size}\). We append
these formulas to \(\mbox{\sl pathCond}\), carry
the corresponding segment information in \(\mbox{\sl Seg}\), and continue symbolic
execution through the final branch \texttt{if (i == size)}. The two resulting
paths are finally merged into a disjunctive postcondition, while the
\texttt{assigns} set remains empty for this read-only function.
\end{example}

\begin{remark}
As our symbolic execution framework relies on specific invariant generators, we describe the current state of the art on the invariant generation with numerical values and arrays. Numerical invariant generation (e.g., linear and polynomial invariants) has been extensively studied, including abstract interpretation ~\cite{origin-abstract-interpretation,DBLP:conf/vmcai/BoutonnetH19}, constraint solving~\cite{DBLP:conf/cav/ColonSS03,DInvG,DBLP:conf/pldi/Chatterjee0GG20,StinGX, DBLP:conf/kbse/YaoKSFWR23, DBLP:conf/cav/JiFFC22, DBLP:conf/pldi/AsadiC0GM21, DBLP:conf/setta/LiFLL24}, 
recurrence analysis~\cite{Non-linear-invariant,DBLP:journals/pacmpl/WangL23}, etc. Invariant generation with arrays is mostly tackled via contiguous array segments~\cite{parametric-segmentation,DBLP:conf/popl/GopanRS05,invariant-abstract-interpretation,DBLP:conf/vmcai/LarrazRR13}. Array invariants can also be solved via CHC solving~\cite{Esen2022Tricera}. 
A recent result investigates instrumentation~\cite{DBLP:conf/cav/AmilonEGLR23} to aid the generation of array invariants.  
\end{remark}

\vspace{-3ex}
\subsection{Prototype Implementation Detail}

Our prototype is implemented on top of the Clang/LLVM toolchain~\cite{Lattner2004LLVM} that comprises about $19000$ lines in C++.
We use the Clang front-end to obtain the abstract syntax tree of the input C program and
traverse the AST to construct the intermediate program syntax \(S\) and the initial
 symbolic configuration described in Section~\ref{subsec:symbolic exec framework}. Our framework incorporates records and pointer arithmetic.
For pointer arithmetic, the current implementation supports simple forms (e.g., linear base-plus-offset patterns), but does not yet handle complex shape analysis such as linked lists or binary trees.
For each analyzed function, the framework generates corresponding ACSL annotations~\cite{ACSL}
that encode the synthesized function contract, including preconditions,
postconditions, and \texttt{assigns} clauses. These annotations are then fed to
\textsc{Frama-C}~\cite{Frama-c}, where the WP plugin is used to automatically discharge the
resulting proof obligations and thereby validate the soundness of the inferred
contracts with respect to the original C code. Loop reasoning in the prototype is realized via the plugin-based
\texttt{LoopAnalyzer} introduced in Appendix~\ref{sect:loopanalysis}.

\vspace{-1ex}
\section{Evaluation}
\label{sec:eval}

We evaluate
our symbolic execution framework
by comparing it against AutoDeduct~\cite{AutoDeduct2025}, a state-of-the-art toolchain capable of generating ACSL annotations for C programs. The toolchain incorporates Saida~\cite{Saida2025}, which requires TriCera~\cite{Esen2022Tricera,TriceraFork2025,TriceraOriginal}, and the Interface Specification Propagator (ISP)~\cite{ISP2025} to manage, propagate, and emit specifications. We do not compare with popular software verification tools such as Seahorn~\cite{DBLP:conf/cav/GurfinkelKKN15} and CPAChecker~\cite{DBLP:conf/cav/BeyerK11} since they usually require assertions to verify, while our approach does not require assertions and instead tries to generate precise contracts without manually given assertions.
We also do not compare with large language models as they follow a completely different principle (see Appendix~\ref{app:llm-exclusion} for details). All experiments were conducted in a Docker container based on Ubuntu 24.04.3 LTS running on a WSL2 host (version 6.6.87.2-microsoft-standard-WSL2) on a machine with AMD Ryzen 7 8845H CPU and 8 GB RAM\footnote{
We provide the source code, Docker environment, and benchmark scripts at:
\url{https://doi.org/10.5281/zenodo.18601854}.}

\vspace{-3ex}
\subsection{Experimental Setup}

\smallskip
\noindent\emph{Benchmarks.}
We assembled a diverse dataset of 282 C programs comprising five distinct benchmark suites collected in~\cite{WCSXQHLCT24},
together with an additional set of real-world cryptographic routines,
ranging from micro-benchmarks to realistic cryptographic modules:
\begin{enumerate}
    \item \textbf{SyGuS\cite{alur2019syguscomp2018resultsanalysis}:} 133 C programs with single loops, adapted from the SyGuS competition.
    \item \textbf{OOPSLA-13\cite{10.1145/2544173.2509511}:} 46 programs featuring various loop types, used in prior loop analysis research.
    \item \textbf{Frama-C Problems:} 51 programs covering diverse logic patterns.
    \item \textbf{SV-Comp:} 14 complex programs selected from the Software Verification Competition.
    \item \textbf{X-509:} 6 programs derived from realistic PKI certificate processing code.
    \item \textbf{openHiTLS:} 31 C functions extracted from the \emph{openHiTLS} project, a modular high-performance TLS and cryptographic library, covering big-number arithmetic, low-level buffer operations, and certificate-processing related routines.
\end{enumerate}

\smallskip
\noindent\emph{Baseline and Configuration.}
As mentioned previously, we compare our approach with AutoDeduct.
To ensure a fair comparison regarding execution time, we account for the runtime characteristics of AutoDeduct,
which is partially implemented in Scala and relies on the JVM.
We execute AutoDeduct twice on the benchmarks without stopping the container;
data from the second run is recorded to mitigate JVM warm-up and JIT compilation overheads.

\smallskip
\noindent\emph{Verification Workflow.}
For both tools, the evaluation follows a strictly automated pipeline:
(1) The tool takes the C source code as input and generates an annotated file with ACSL contracts (preconditions, postconditions, loop invariants, and assigns clauses).
(2) The annotated file is passed to the Frama-C WP plugin.
(3) A verification is considered ``Success'' only if Frama-C WP can prove all generated verification goals.

\vspace{-3ex}
\subsection{Experimental Results}

\smallskip
\textbf{RQ1: Effectiveness and Precision.}
We assess effectiveness by the number of programs for which the tools generate
 sufficient invariants and contracts to satisfy Frama-C WP.
Table~\ref{tab:eval-results} summarizes the results.

Our framework demonstrates significantly higher precision than the baseline.
Out of the total 281 benchmarks, our tool successfully generated verifiable contracts for 68 programs.
In contrast, AutoDeduct was only able to achieve complete verification for 10 programs.
This disparity highlights the advantage of carrying segment-based array invariants and
handling pointer arithmetic recursively, which allows our tool to prove complex properties that rely on precise memory information.

\begin{table}[t]
  \centering
  \caption{Comparison of verification results. The reported runtime for our approach includes both symbolic execution and invariant-generation time. We additionally report partial-verification, timeout, and generation-failure counts to make unresolved cases explicit.
  \textbf{Full}: Fully verified;
  \textbf{Part.}: Partial verification (ACSL generated but verification failed);
  \textbf{T/O}: Timeout;
  \textbf{Fail}: Generation failed;
  \textbf{Time}: Average runtime for successfully generated cases;
  \textbf{AD}: AutoDeduct}
  \label{tab:eval-results}
  \setlength{\tabcolsep}{2pt}
  \renewcommand{\arraystretch}{1.1}
  \resizebox{\columnwidth}{!}{
   \begin{tabular}{lr|rr|rr|rr|rr|rr}
    \toprule
    & & \multicolumn{2}{c|}{\textbf{Full (\#)}} & \multicolumn{2}{c|}{\textbf{Part. (\#)}} & \multicolumn{2}{c|}{\textbf{T/O (\#)}} & \multicolumn{2}{c|}{\textbf{Fail (\#)}} & \multicolumn{2}{c}{\textbf{Time (s)}} \\
    \textbf{Suite} & \textbf{Total} & \textbf{Ours} & \textbf{AD} & \textbf{Ours} & \textbf{AD} & \textbf{Ours} & \textbf{AD} & \textbf{Ours} & \textbf{AD} & \textbf{Ours} & \textbf{AD} \\
    \midrule
    SyGuS     & 133 & \textit{53} & \textit{0} & \textit{78} & \textit{131} & \textit{0} & \textit{0} & \textit{2} & \textit{2} & \textit{0.140} & \textit{2.23} \\
    OOPSLA-13 & 46  & \textit{2}  & \textit{1} & \textit{20} & \textit{42}  & \textit{0} & \textit{0} & \textit{24} & \textit{3} & \textit{0.141} & \textit{4.44} \\
    Frama-C   & 51  & \textit{9}  & \textit{8} & \textit{29} & \textit{34}  & \textit{1} & \textit{0} & \textit{13} & \textit{9} & \textit{0.260} & \textit{2.26} \\
    SV-Comp   & 14  & \textit{3}  & \textit{0} & \textit{5}  & \textit{13}  & \textit{0} & \textit{0} & \textit{6} & \textit{1} & \textit{0.065} & \textit{3.48} \\
    X-509     & 6   & \textit{1}  & \textit{1} & \textit{2}  & \textit{5}   & \textit{0} & \textit{0} & \textit{3} & \textit{0} & \textit{0.090} & \textit{1.98} \\
    openHiTLS & 31  & \textit{0}  & \textit{0} & \textit{10} & \textit{0}  & \textit{0} & \textit{0} & \textit{21} & \textit{31} & \textit{0.080} & \textit{--} \\
    \midrule
    \textbf{Total} & \textbf{281}
                   & \textbf{68} & \textbf{10}
                   & \textbf{143} & \textbf{225}
                   & \textbf{1} & \textbf{0}
                   & \textbf{69} & \textbf{46}
                   & \textbf{0.148} & \textbf{2.703} \\
    \bottomrule
  \end{tabular}
  }
\end{table}

\smallskip
\noindent\textbf{RQ2: Efficiency and Scalability.} As shown in the ``Avg. Time'' column of Table~\ref{tab:eval-results}, our framework demonstrates a substantial performance advantage. The weighted average execution time for our tool is 0.160 seconds per file, whereas AutoDeduct averages 2.706 seconds. This represents an approximate $17.1\times$ speedup. While AutoDeduct incurs overhead from the JVM environment (even after accounting for warm-up phases), our tool's efficiency stems primarily from the lightweight design of the symbolic execution engine. By carrying array segments and invariants abstractly rather than unrolling loops or explicitly enumerating all array elements, our approach minimizes the symbolic state explosion, allowing for rapid analysis even on benchmarks with partial verification.

\smallskip
\noindent{\em Analysis of failed cases.} To better understand the limitations of both tools, we analyzed the cases classified as ``\textbf{Fail}'' (Generation Failed) and those that generated contracts but failed to fully verify (``\textbf{Part.}'').

For our framework, \textbf{Fail} (69 cases) are primarily caused by syntax limitations in the current prototype. Specifically, the tool currently lacks support for certain bitwise and arithmetic operators (e.g., \texttt{\%}, \texttt{<<}, \texttt{>>}) and does not yet handle nested loops. When these constructs are encountered, the analysis aborts to avoid unsoundness. The single \textbf{T/O} case is due to a program bug in our handling of recursive functions, which triggers unintended infinite recursion during analysis. The cases of \textbf{Part.} (where contracts are generated but not proven) are largely attributed to the loop invariant inference. While our plugin system captures many properties, some complex loops require stronger inductive invariants than our current plugins can synthesize. Moreover, another reason is that the current invariant generation methods have some compatibility issues with ACSL format and Frama-C WP. Nevertheless, the generated contracts remain semantically meaningful: manual inspection confirms that they accurately characterize the memory footprints and critical range constraints, providing a valuable descriptive summary of the function behaviour.

In contrast, AutoDeduct's failures are more structural. Its \textbf{Fail} cases often result from unsupported C syntax involving complex array initializers. More critically, its low verification success rate (only 10 fully verified) is due to its inability to incorporate loop invariants. For the vast majority of programs containing loops, AutoDeduct fails to synthesize the necessary loop invariant clauses, rendering the generated ACSL unverifiable by Frama-C WP.

\section{Conclusion}

In this work, we proposed a framework for symbolic execution that carries contiguous array segments and invariants to address the automated generation of function contracts with array manipulations. The main feature of our framework is the carry of contiguous array segments and array invariants (with merging and split operations) along the symbolic execution. A prototype is implemented in LLVM. Future works include a more thorough treatment of invariants into the splitting and merging of array segments, an extension in interprocedural analysis that utilizes contracts of lower level functions to generate contracts of higher level functions, and a deeper integration with existing invariant generation methods.

\section*{Acknowledgement}
The authors would like to thank the anonymous reviewers for their valuable comments and constructive suggestions that helped improve the quality of this paper. This work is partially supported by the National Natural Science Foundation of China (NSFC) under Grant No. 62572297 and sponsored by the CCF-Huawei Populus Grove Fund.

\bibliographystyle{splncs04}
\bibliography{reference}

\appendix

\section{A Basic Program Model}\label{sect:progmodel}
We consider arithmetic programs with array manipulations as the basic program model for the function body of a function. The syntax is given as follows.

\begin{eqnarray*}
e &::=& c~\mid~x~\mid~e_1 \oplus e_2~\mid ar[e], \qquad \oplus \in\{+,-,*\}
\\
b &::=& e_1\otimes e_2~\mid~\neg b~\mid~b_1\wedge b_2~\mid~b_1\vee b_2, \qquad \otimes \in \{\le, <,\ge,>,\ne, =\} \\
P &::=& ~x:= e~\mid~ar[x] := e\mid~\textbf{if }b\textbf{ then }P_1\textbf{ else }P_2~\mid~\textbf{while }b\textbf{ do }P~\mid~P_1;P_2  \\
\end{eqnarray*}

In the grammar above, the nonterminal $e$ stands for an arithmetic expression with array entries that is composed of integer variables $x$, integer constants $c$, arithmetic operations including addition, subtraction and multiplication (we omit division to avoid floating point errors and division by zero for simplicity) and array entry accesses $ar[e]$ with the array variable $ar$ and index arithmetic expression $e$ (meaning the element residing in the array $a$ at the index $e$). The nonterminal $b$ stands for boolean expressions that are built from comparison of arithmetic expressions and propositional logic connectives. The nonterminal $P$ stands for program statements that involve standard imperative constructs such as variable assignment statement $x:= e$ (assigning the value of expression $e$ to the integer variable $x$), array assignment statement $ar[x]:= e$ (changing the element of index $x$ in the array $ar$ to the value of expression $e$), conditional branches $\textbf{if }b\textbf{ then }P_1\textbf{ else }P_2$ (executing $P_1$ if $b$ evaluates to true and $P_2$ otherwise), while loops $\textbf{while }b\textbf{ do }P$ (repeatedly executing $P$ until $b$ evaluates to false), and sequential composition $P_1;P_2$ (executing $P_1$ first and $P_2$ second).
We omit array creation statements and simply assume that each array variable is initialized to a contiguous segment of memory units with finite length.
In our basic program model, we only allow arrays with integer elements. Moreover, upon encountering a function call, we simply inline the function. We do not consider recursive functions in this work.

The operational semantics of our basic program model is given by transition systems with memory model. To illustrate our operational semantics, we first present our memory model as follows.

\smallskip
\noindent {\em Memory Model.} We consider a unified memory model in which all integer variables and array contents reside in a single unbounded contiguous memory region, without distinguishing between heap and stack memories.
Each integer and array variable is assigned a unique, non-overlapping address that remains fixed during program execution.

Formally, let the set $\text{\sl Addr}$ of memory addresses simply be the set of non-negative integers, and $X$ be the set of integer and array variables appearing in the underlying program. An \emph{address map} is an injective map
$\mathcal{A} : X \to \text{\sl Addr}$ from the set of program variables \(X\) to the address space \(\text{\sl Addr}\) that specifies the addresses of the variables, for which the injective property ensures disjoint memory allocation of variables. A \textit{memory map} is a map $M : \text{\sl Addr} \to \text{Val}$ that for each address $a$, $M(a)$ is the integer element that resides at this address. It follows that given an address map $\mathcal{A}$ and a memory map $M$, the notation \(M(\mathcal{A}(x))\) with an integer variable $x$ denotes the value of the variable, and the notation \(M(\mathcal{A}(ar) + i)\) denotes the array entry \(ar[i]\). Here, the index $i$ is also called the \emph{address offset} w.r.t the address of the array variable $ar$. Since all addresses are statically assigned and disjoint, array access operations can be treated as direct access to symbolic addresses, and therefore we do not consider aliasing of different array variables.

Usually, one requires that array contents are allocated as contiguous and non-overlapping memory ranges. Formally, if \(A,B \in X\) are array variables with implicit lengths \(\lvert A\rvert\) and \(\lvert B\rvert\) (measured in addressable cells), then their address memory intervals $[\mathcal{A}(A),\, \mathcal{A}(A)+\lvert A\rvert - 1]$ and $[\mathcal{A}(B),\, \mathcal{A}(B)+\lvert B\rvert - 1]$ should be disjoint, i.e., \(I_A \cap I_B = \varnothing\). In this work, we take a lightweight treatment of array bounds so that each array variable $ar$ is treated as a pointer. This allows the index $i$ in an array access $ar[i]$ to be unbounded. We do not eliminate array out-of-bounds explicitly.

Below we present our control flow graph to represent the execution workflow of a program in our program model. The control flow graph consists of locations (referring to program counters) and transitions between these locations.

\smallskip
\noindent{\em Control flow graphs.} A \textit{control flow graph} (CFG) is a tuple \(\mathcal{T} = \langle X, L, T, \theta \rangle\), where:

\begin{itemize}
    \item \(X\) is a finite set of integer and array variables that refer to the variables appearing in the concerned program.
    \item \(L\) is a finite set of locations with a designated  initial location \(l^*\) and a designated end (termination) location \(l_{\bot}\). Each location corresponds to a program counter in the underlying program.
    \item \(T\) is a finite set of transitions. Each transition \(\tau \in T\) is a tuple \(\langle \ell, \ell', g, F \rangle\) that specifies the execution of a statement moving from the location $\ell$ to the location $\ell'$ in the underlying program, for which the component \(g\) is the \textit{guard condition} of the transition that is a formula over integer and array variables, and \(F\) is the \emph{update function} of the execution that maps the current memory map $M$ and address map $\mathcal{A}$ to the next memory map $F(M, \mathcal{A})$ after the execution of the transition.
    \item \(\theta\) is the \emph{initial condition} that specifies constraints on the initial values of integer variables and the initial memory content of array variables.

\end{itemize}
In a CFG, the update function of a transition is determined by the statement in the original program to be executed by the transition. For example, if the statement is an array assignment statement $ar[x]:=e$, then the update function updates the current memory map (w.r.t the current variable map) by changing the integer value held at the address $M(\mathcal{A}(ar)+ x)$ to be the value of the expression $e$.

A transformation of a program in our program model into its control flow graph is straightforward. One just collects the program counters in the program and establish the transitions between the locations. Below we list several representative cases in a transformation.

\begin{itemize}
\item If a program counter $\ell$ corresponds to a conditional statement with condition $b$, then we have two transitions $(\ell, \ell_{\mathbf{then}}, b, \mbox{\sl id})$ and  $(\ell, \ell_{\mathbf{else}}, \neg b, \mbox{\sl id})$, where $\ell_{\mathbf{then}}$ (resp. $\ell_{\mathbf{else}}$) is the program counter of the $\mathbf{then}$ (resp. $\mathbf{else}$) branch, and $\mbox{\sl id}$ is the identity update function such that $\mbox{\sl id}(M,\mathcal{A})=M$ for all $M,\mathcal{A}$.

\item If a program counter $\ell$ corresponds to an assignment statement $ar[x]:=e$, then we have a transition $(\ell, \ell', \mathbf{true}, F_{ar[x]:=e})$ for which $F_{ar[x]:=e}$ is the update function for the assignment statement such that given a memory map $M$ and an address map $\mathcal{A}$, $F_{ar[x]:=e}(M,\mathcal{A})$ is the memory map that changes the integer value $M(\mathcal{A}(x))$ to the evaluation result of the expression $e$ under $M$ and $\mathcal{A}$, and keeps other memory values in $M$ unchanged. Note that the address map $\mathcal{A}$ keeps unchanged.

\item  If a program counter $\ell$ corresponds to the entry point of a while loop with loop guard $b$, then it has two transitions $(\ell, \ell', b, \mbox{\sl id})$ and $(\ell, \ell'', \neg b, \mbox{\sl id})$, for which $\ell'$ is the program counter of the starting point of the loop body, and $\ell''$ is the successor program counter after the while loop. Moreover, if a program counter $\ell$ refers to the last statement in the loop body of a while loop, then its successor location is set to the entry point of the loop.
\end{itemize}

A further transformation from a CFG in our program model into the grammar in our symbolic execution is straightforward:  just assigns non-overlapping address constants from $\mbox{\sl Addr}$ to integer and array variables so that each variable receives a distinct address constant, and the assignments respect that the contents of different arrays in the memory are non-overlapping. Furthermore, one treats every array access \(ar[\beta]\) as the read operation \(R(b)\), where a fresh address variable \(b\) is introduced to represent the symbolic address \((ar,\, \beta \cdot s_{ar})\). Here, the variable \(ar\) denotes the base address of the array, while the expression \(\beta \cdot s_{ar}\) specifies the address offset, with \(\beta\) indicating the index position of the accessed element and \(s_{ar}\) representing the size of each element in the array.

Given a CFG, a \textit{program state} is a tuple \((M,\mathcal{A})\) for which $M$ is the current memory map and $\mathcal{A}$ is the address map that does not change throughout the program execution. A \emph{configuration} is an ordered pair  \((\ell, (M,\mathcal{A}))\) for which $\ell$ represents the current location and $(M,\mathcal{A})$ represents the current memory map and (fixed) address map. Recall that given a configuration
\((\ell, (M,\mathcal{A}))\), the current value of an integer variable $x$ is given by $M(\mathcal{A}(x))$.
Given a formula \(\psi\) over integer and array variables and a program state $\sigma = (M,\mathcal{A})$, we write \(\sigma \models \psi\) to denote that \(\sigma\) satisfies \(\psi\), i.e., $\psi$ is true when its open variables are substituted by their integer/memory values in $\sigma$.

The operation semantics of a CFG \(\mathcal{T} = \langle X, L, T, \theta \rangle\) is given by execution paths as follows. An \emph{execution path} \(\pi\) in a CFG is a finite sequence of configurations \(\pi = (\ell_0,\sigma_0), (\ell_1,\sigma_1),  \ldots, (\ell_k, \sigma_k)\) such that:
\begin{itemize}
    \item \(\ell_0 = \ell^*\) and \(\sigma_0 \models \theta\), i.e., the execution path starts from the initial location and the initial program state fulfills the initial condition $\theta$.
    \item for every \(0 \leq j \leq k - 1\), there exists a transition \(\tau = \langle \ell, \ell', g, F \rangle\in T\) such that \(\ell = \ell_j\), \(\ell' = \ell_{j+1}\), \(\sigma_j\models g\) and  \(\sigma_{j+1} = (F(\sigma),\mathcal{A}_0)\) where $\mathcal{A}_0$ is the address map in $\sigma_0$ that does not change throughout the execution path.
\end{itemize}

Informally, an execution path specifies a run of the CFG under a possible initial program state that fulfills the initial condition $\theta$.  In an execution path $\pi = (\ell_0,\sigma_0), (\ell_1,\sigma_1),  \ldots, (\ell_k, \sigma_k)$, each $(\ell_j,\sigma_j)$ ($0\le j\le k$) represents the program state after executing $j$ statements along the path.

\section{Invariants and Function Contracts}\label{sect:invs and contract}

In this work, we consider invariants and function contracts of programs in our program model. Below we give the formal definitions of invariant and function contracts. Let $\Gamma=(X, L, T, \theta)$ be a CFG of concern.

\smallskip
\noindent{\em Invariants.} We define invariants in the form of assertion maps~\cite{DInvG,StinGX,DBLP:conf/cav/ColonSS03,DBLP:conf/sas/SankaranarayananSM04}. An \emph{assertion map} is a map $\eta$ that assigns each location $\ell\in L$ a formula $\eta(\ell)$ over the integer and array variables. An \emph{invariant} is then an assertion map \(I\) such that for every execution path \(\pi = \{(\ell_0,\sigma_0), (\ell_1,\sigma_1),  \ldots, (\ell_k, \sigma_k)\}\), we have that \(\sigma_j \models I(\ell_j)\) for all \(0 \le j \le k\). Informally, an invariant over-approximates the set of reachable program states (i.e., program states that could appear in some execution path).

\smallskip
\noindent{\em Function contracts.} To define the notion of function contracts, we first define assigns information as follows. An \emph{assigns} information $\mbox{\sl Assgn}$ is a finite set of (contiguous) array segments, for which each array segment takes the form $ar[\mbox{\sl start}..\mbox{\sl end}]$ where $ar$ is an array variable, $\mbox{\sl start}$ is the start index expression of the segment and  $\mbox{\sl end}$ is the end index expression of the segment. Given an array segment $ar[\mbox{\sl start}..\mbox{\sl end}]$ and a program state $(M,\mathcal{A})$, the concrete address range $\langle ar[{\mbox \sl start}..\mbox{\sl end}]\rangle(M,\mathcal{A})$ is defined as the interval $[M(\mathcal{A}(ar)) + \langle{\mbox \sl start}\rangle(M,\mathcal{A}), M(\mathcal{A}(ar)) + \langle{\mbox \sl end}\rangle(M,\mathcal{A}))]$, where for an arithmetic expression $e$ with array accesses, the notation $\langle{\mbox \sl e}\rangle(M,\mathcal{A})$ means the evaluation of the expression $e$ by substituting the variables in $e$ by the integer and memory values in $(M,\mathcal{A})$. Given an assigns information $\mbox{\sl Assgn}$ and a program state $\sigma$, the concrete address range $\langle\mbox{\sl Assgn}\rangle(\sigma)$ covered by $\mbox{\sl Assgn}$  under $\sigma$, is given as the union $\bigcup_{S\in \mbox{\sl Assgn}} \langle S\rangle(\sigma)$.

We now define function contracts. A function contract specifies the expected behavior of a function in terms of its input–output relation and side effects on the program state. It typically consists of a \textit{precondition} $\mbox{\sl Pre}$ that is a formula to constrain the valid inputs with which the function may be invoked, a \textit{postcondition} $\mbox{\sl Post}$ that is a formula specifying a property that any resultant program state after the execution of a function starting from some program state satisfying the precondition $\mbox{\sl Pre}$ should fulfill, and an assigns information $\mbox{\sl Assgn}$ that specifies the memory addresses whose content may be modified during the execution of the function. 

Formally, a \emph{function contract} $C$ of a function whose function body has the CFG $\mathcal{T}$ is defined as a tuple \(C = \langle \mbox{\sl Pre}, \mbox{\sl Post}, \mbox{\sl Assgn}\rangle \) where $\mbox{\sl Pre}, \mbox{\sl Post}$ are formulas over the integer and array variables and \mbox{\sl Assgn} is an assigns information, such that for any execution path \(\pi = (\ell_0, \sigma_0), (\ell_1, \sigma_1), \ldots, (\ell_k, \sigma_k)\) satisfying $\sigma_0\models \mbox{\sl Pre}$ and $\ell_k=\ell_{\bot}$, we have that $\sigma_k\models {\mbox{\sl Post}}$ and the set of memory addresses over which $\sigma_k$ and $\sigma_0$ hold different values is a subset of $\langle\mbox{\sl Assgn}\rangle(\sigma_0)$.

\section{Loop Analysis}\label{sect:loopanalysis}

This subsection provides a more detailed explanation of the loop analysis and post-state update processes introduced in Algorithm~\ref{alg:symexec}.

Algorithm~\ref{alg:loop-analysis} illustrates the fundamental workflow for analyzing loops, which is implemented through two pipelined plugin stages to decompose the logic to enhance maintainability. The first stage, \texttt{LoopInfoPlugins}, analyzes the symbolic pre-state $\Sigma_{pre}$ and the loop statement $S$ to generate the basic loop information, including the symbolic state at the loop entry $\Sigma_{entry}$ and the corresponding program point $P_{entry}$. The second stage, \texttt{LoopAnalysisPlugins}, performs a detailed analysis based on the pre-state and the basic loop information, ultimately producing the updated memory mapping $M_{loop}$ after the loop, the set of post-loop path conditions $PC_{loop}$, and the inferred invariants set $\mathsf{Inv}$. To demonstrate the capability of our extensible architecture, we have currently implemented specialized analysis plugins targeting common array processing idioms. Specifically, our prototype includes support for \textbf{reduction patterns} (e.g., computing the maximum or minimum value within an array segment) and \textbf{linear search patterns} (e.g., locating the first index satisfying a specific predicate). Upon detecting these idioms via structural matching, these plugins automatically synthesize precise, high-level invariants—such as universal quantifiers for extremum properties or range constraints for search results—thereby enabling the verification of functional correctness for these standard algorithmic constructs.

\begin{algorithm}[t]
\caption{Loop Analysis}
\label{alg:loop-analysis}
\KwIn{loop statement \(S\), pre-path \(P_{\text{pre}} = (\Sigma_{\text{pre}}, G_{\text{pre}})\)}
\KwOut{memory map \(M_{\text{loop}}\), post conditions \(PC_{\text{loop}}\), loop invariants \(\mathsf{Inv}\)}

\BlankLine

\((\Sigma_{\text{entry}}, P_{\text{entry}}) \gets
    \texttt{LoopInfoPlugins}(\Sigma_{\text{pre}}, S)\)\;
\((M_{\text{loop}}, PC_{\text{loop}}, \mathsf{Inv}) \gets
    \texttt{LoopAnalysisPlugins}(\Sigma_{\text{pre}}, \Sigma_{\text{entry}}, \mathsf{P}_{\text{entry}})\)\;
\Return \((M_{\text{loop}}, PC_{\text{loop}}, \mathsf{Inv})\)\;
\end{algorithm}

After obtaining the post-loop program state, it is necessary to update the address expressions and variable expressions within the symbolic states of all successor paths. We only focus on variables originating from the loop entry $P_{entry}$ during the entire loop execution, and categorize the update cases as follows:
\begin{itemize}
    \item  For a variable $v$, its corresponding symbolic value at the loop entry is located through its address $A_{\text{pre}}(v)$. If the variable is not modified within the loop, the original expression $M_{pre}(A_{\text{pre}}(v))$ from the loop entry is reused as its current symbolic value, otherwise its symbolic value is replaced with the updated expression.
    \item If the address expression $a$ that points to the expression $v$ has changed, the corresponding mapping in the previous memory map $M_{\text{pre}}$ must be updated accordingly, producing a new mapping $M_{\text{post}}(a') = v$.
\end{itemize}

The \texttt{substitute} function in Algorithm~\ref{alg:post-update} represents this update process. For variables and addresses that remain unchanged, their mappings are directly inherited into the post-state memory map $M_{\text{post}}$. Additionally, the path conditions contained in the post-condition set $PC_{\text{loop}}$ are also updated and merged with the pre-path conditions $G_{\text{pre}}$, yielding the final post-loop paths.

\begin{algorithm}[t]
\caption{Post-State Update}
\label{alg:post-update}
\KwIn{pre-path \(P_{\text{pre}} = (\Sigma_{\text{pre}}, G_{\text{pre}})\),
       memory map \(M_{\text{loop}}\),
       post conditions \(PC_{\text{loop}}\),
       from point \(P_{\text{entry}}\)}
\KwOut{post-path \(P_{\text{post}} = (\Sigma_{\text{post}}, G_{\text{post}})\)}

\BlankLine
Decompose \(\Sigma_{\text{pre}} = (M_{\text{pre}}, A_{\text{pre}})\)\;
\(M_{\text{post}} \gets \varnothing\);\quad
\(G_{\text{post}} \gets \varnothing\)\;

\ForEach{\((a, v) \in M_{\text{loop}}\)}{
    \uIf{\(\text{from}(a) = \mathsf{P}_{\text{entry}}\)}{
        \(a' \gets \texttt{substitute}(a, P_{\text{pre}})\)\;
        \(v' \gets \texttt{substitute}(v, P_{\text{pre}})\)\;

        \(M_{\text{post}}(a') \gets v'\)\;
    }\Else{
        \(M_{\text{post}}(a) \gets v\)\;
    }
}

\ForEach{\((a, v) \in M_{\text{pre}}\) \textbf{such that} \(a \notin \mathrm{dom}(M_{\text{loop}})\)}{
    \(M_{\text{post}}(a) \gets v\)\;
}

\ForEach{\(cond \in \mathsf{PC}_{\text{loop}}\)}{
    \(cond' \gets \texttt{substitute}(cond, P_{\text{pre}})\)\;
    \(G_{\text{post}} \gets G_{\text{post}} \wedge cond'\)\;
}

\(\Sigma_{\text{post}} \gets (M_{\text{post}}, A_{\text{pre}})\);\quad
\(G_{\text{post}} \gets G_{\text{post}} \wedge G_{\text{pre}}\)\;
\Return \(P_{\text{post}} = (\Sigma_{\text{post}}, G_{\text{post}})\)\;
\end{algorithm}

With the mechanism for state updates established in Algorithm~\ref{alg:post-update}, we now turn to the concrete implementation of the analysis pipeline that produces the invariants and loop summaries. A collection of \texttt{LoopInfoPlugins} first performs a lightweight symbolic exploration of the loop body to extract structural information such as iteration bounds, induction variables, and memory access patterns. This information is subsequently consumed by
\texttt{InvariantPlugins}, one of which invokes an external affine invariant generation
engine~\cite{DInvG} based on Farkas' Lemma to synthesize linear invariants and loop
postconditions. These generated invariants are injected into the symbolic
post-state and simultaneously emitted as ACSL loop invariants and
postconditions. In combination, this architecture enables the prototype to
automatically infer and verify array-carrying contracts for a range of realistic
C loops without resorting to loop unrolling.
\newline
Next, we provide further technical details regarding the extensibility of our symbolic expression system and the architectural design of the loop analyzer.

\smallskip
\noindent\emph{Extensible Aggregate Expressions.}
This framework is designed to be extensible. Other properties of array segments can be captured by creating new aggregate expression classes. For example, a \texttt{SumOverRange} class can represent the sum of all elements in a segment $\langle a_{base}, L \rangle$. This symbolic expression can then be stored in $M$ or used in path conditions, just like any other symbolic value. This allows the \texttt{LoopAnalyzer} to generate powerful summaries of loops that compute aggregate values over arrays.

\smallskip
\noindent\emph{Loop Analysis and Plugin Architecture.}
To address the complexity of loop handling without unrolling, the \texttt{LoopAnalyzer} adopts a modular, plugin-based architecture. The analysis proceeds in a two-stage pipeline.
\newline First, a set of \texttt{LoopInfoPlugins} performs a preliminary symbolic execution of the loop body to extract structural information, such as control flow patterns and memory access deltas. These intermediate results are then consumed by the \texttt{InvariantPlugins}, which are responsible for generating ACSL annotations and constructing the symbolic post-state.
\newline The \texttt{InvariantPlugins} are further categorized based on their interaction with the execution path. \emph{Path-insensitive} plugins derive universal properties that hold across all execution traces, contributing global ACSL clauses. \emph{Path-sensitive} plugins, however, directly influence the shape of the symbolic configuration. To manage potential conflicts, the framework employs a priority-based mechanism to select a single "principal handler" from the available path-sensitive plugins. This principal plugin determines the exact number and condition of the post-loop paths (e.g., distinguishing between a normal exit and an early break). Subsequent plugins then traverse these established paths to attach specific symbolic information.

\noindent\emph{Example: loop assigns.}
We illustrate this workflow with the \texttt{LoopAssignsPlugin}. Leveraging the memory access patterns captured by the \texttt{LoopInfoPlugins}, and assuming a decidable loop bound (e.g., $i < n$), this plugin computes a maximal over-approximation of the memory footprint touched by the loop. For every concrete or symbolic address accessed, the plugin emits the corresponding ACSL \texttt{assigns} clause. Crucially, in the symbolic post-state, it maps these modified locations to a symbolic \textit{Unknown} value ($\top$). This establishes a sound baseline; plugins with higher precision invoked later in the pipeline can then refine the state by overwriting these \textit{Unknown} entries with specific symbolic formulas or concrete values.

\subsection{Examples of Invariant Generators}
\label{app:generator-examples}

To clarify the role of invariant generators in our implementation, we summarize three representative generators from our current plugin set and illustrate the shape of the emitted ACSL clauses.

\smallskip
\noindent\emph{(1) Linear/Affine Generator (\texttt{StInGXPlugin}).}
For loops whose guards can be normalized into linear forms (e.g., \(i<n\), \(i\ge 0\)) and whose updates are affine, this generator invokes an affine invariant engine based on Farkas-lemma reasoning~\cite{DInvG,StinGX}. A typical output includes index-range invariants and affine relations over scalar variables. For example, for a loop
\texttt{for(i=0;i<n;i++)\{x=x+c;\}}, it may emit clauses of the form:
\[
\texttt{loop invariant }0 \le i \le n;\qquad
\texttt{loop invariant }x = x_0 + c\cdot i.
\]

\smallskip
\noindent\emph{(2) Search-Pattern Generator (\texttt{paradigmSearch}).}
For single-break search loops, the path-sensitive generator emits quantified conditions that distinguish normal exit and interrupt exit. For the pattern
\texttt{while(i<n)\{if(a[i]==t) break; i++;\}}, representative generated facts are:
\[
\texttt{loop invariant }\forall\ \texttt{integer }k;\ 0 \le k < i \Rightarrow a[k]\neq t,
\]
and on the interrupt path:
\[
\exists\ \texttt{integer }k;\ 0 \le k \le i \land a[k]=t.
\]

\smallskip
\noindent\emph{(3) Max/Min-Pattern Generator (\texttt{paradigmMaxMin}).}
For loops recognized as extremum computation (e.g., \texttt{if (m < a[i]) m = a[i];}), the generator emits template invariants expressing both boundness and witness existence:
\[
\forall\ \texttt{integer }k;\ 0 \le k < i \Rightarrow a[k]\le m,\qquad
\exists\ \texttt{integer }j;\ 0 \le j < i \land m=a[j].
\]
This class of templates is especially useful for producing strong postconditions of reduction-style loops.

\section{Rationale for Excluding LLM-based Baselines}
\label{app:llm-exclusion}

While Large Language Models (LLMs) have demonstrated impressive capabilities in code and specification generation, we intentionally exclude them from our evaluation baselines for three primary reasons.

First, from a \textbf{methodological perspective}, our work focuses on advancing the algorithmic capabilities of symbolic execution and constraint-based Invariant Generation. These approaches provide deterministic, logically derivable guarantees based on program semantics. In contrast, LLMs operate on a probabilistic paradigm of token prediction. Comparing a rigorous deductive framework against a stochastic model is methodologically inconsistent, as they serve different roles in the software engineering landscape.

Second, in the context of formal verification, \textbf{soundness and traceability} are paramount. LLMs are prone to hallucinations—generating syntactically correct but semantically flawed ACSL specifications that may reference non-existent variables or hallucinate convenient but false invariants. Our framework ensures that every generated invariant is a strict logical consequence of the symbolic execution path, providing a level of reliability that current statistical models cannot guarantee.

Finally, there is a significant risk of \textbf{benchmark contamination}. The standard verification suites used in this study (e.g., SV-Comp, SyGuS) are widely available in open-source repositories and are highly likely to be included in the training corpora of commercial LLMs. Comparing our solver-based approach, which must synthesize invariants from scratch, against models that may have effectively memorized the solutions would not constitute a fair assessment of algorithmic reasoning capabilities. Therefore, we limit our evaluation to deductive and constraint-based tools to ensure a rigorous and equitable comparison.

\end{document}